# A General Exact Closed-Form Solution for Nonlinear Differential Equation of Pendulum

Mohammad Asadi Dalir [1]

[1] Mechanical Engineering Department, Bu-Ali Sina University, Hamedan, Iran

m.asadi@alumni.basu.ac.ir

**Abstract.** In the present paper, the nonlinear differential equation of pendulum is investigated to find an exact closed form solution, satisfying governing equation as well as initial conditions. The new concepts used in the suggested method are introduced. Regarding the fact that the governing equation for any arbitrary system represents its inherent properties, it is shown that the nonlinear term causes that the system to have a variable identity. Hence, the original function is included as a variable in the solution to can take into account the variation of governing equation. To find the exact closed form solution, the variation of the nonlinear differential equation tends to zero, where in this case the system with a local linear differential equation has a definite identity with a definite local answer. It is shown that the general answer is an arbitrary curve on a surface, a newly developed concept known as super function, and different initial conditions give different curves as particular solutions. The comparison of the results with those of finite difference shows an exact agreement for any arbitrary amplitude and initial conditions.

**Key words**: Nonlinear differential equation, exact closed form solution, super function, variable identity

## 1. Definitions

### 1.1 Constant differential equation

A differential equation in the form of $L(\ddot{x}, \dot{x}, x, a_j)$ in which $x = x(t)$ and $a_j$ ($j = 1,2$) is the constant coefficient of differential equation, is known as constant differential equation with a constant identity in all times and amplitudes of $x$. The governing equation of a linear mass-spring system $\ddot{x} + \omega^2 x = 0$ is an example of a constant differential equation.

### 1.2 Function with constant identity

A definite function with a definite identity with the form $x(t, a_j)$ in which by knowing initial condition its value and behavior is determined for all points is a function with constant identity which is called constant function here, for abbreviation. A constant function can be the answer of a constant differential equation whose most important property in this definition is that the points in an infinitesimal neighbourhood satisfy the slope definition for a function below:

$$\frac{dx(t, a_j)}{dt} = \lim_{\Delta t \to 0} \frac{x(t + \Delta t, a_j) - x(t, a_j)}{\Delta t} \qquad (1)$$

The function $x(t, \omega) = \sin(\omega t)$ is an example of a constant function.

### 1.3 Variable differential equation

A differential equation in the form $L(\ddot{x}, \dot{x}, y(x), a_j)$, where $y(x)$ is a known function of $x$, is called a variable differential equation in which the coefficients of differential equation can vary with $x$. As a result, the coefficients of differential equation are constant for a determined $x$, but vary by amplitude. Thus, the differential equation has a variable identity. The pendulum equation $\ddot{x} + \omega^2 \sin x = 0$ is an example of a variable differential equation.

### 1.4 Super function

A mathematical concept existing by the form $x\left(t^i, a_j(x^*)\right)$ is called a super function, where, $x^*$ is a continuous variable and the local time $t^i$ represents several scales of time for $i = 0,1,2,...$ . Given the presence of the continuous variable $x^*$, a super function can be reduced to a constant function $x\left(t^1, a_j(k_1)\right)$ for $x^* = k_1$ (i.e. $k_1$ = cte) and the corresponding time scale $t^1$. Here the author emphasizes that a super function is neither a one- variable function nor a multi-variable function. Regarding the fact that each constant function has definite properties, the second variable $x^*$ in a super function is inserted to obtain desired properties of several local constant functions by their arbitrary combination. Thus, $x^*$ varies in such a way that the under-consideration super function has the required behavior. Consider a super function that has the properties of $x\left(t^1{}_0, a_j(k_1)\right)$, $x\left(t^2{}_0, a_j(k_2)\right)$ and $x\left(t^3{}_0, a_j(k_3)\right)$ in the times $t - \Delta t$, $t$ and $t + \Delta t$ ,respectively. In the recent equation $t^i{}_0$ is a determined moment of $t^i$ when the behavior of super function is according to that of local constant function $x\left(t^i, a_j(x^*)\right)$ which we call, active local constant function. Therefore, the slope of a super function is defined according to the following equation:

$$\frac{dx\left(t^i, a_j(x^*)\right)}{dt} = \lim_{\Delta t \to 0} \frac{x\left(t^{i+1}{}_0, a_j(x^* + \Delta x^*)\right) - x\left(t^i{}_0, a_j(x^*)\right)}{\Delta t} \tag{2}$$

in which, when $\Delta t \to 0$, we have $\Delta x^* \to 0$ as well. This equation tells us that the amount of answer and its local derivatives in the moment $t$ is $x\left(t^i, a_j(x^*)\right)$ and $d^n x\left(t^i, a_j(x^*)\right)/dt^{i^n}$, respectively at $t^i = t^i{}_0$ and a moment later in $t + \Delta t$ the amount of these quantities is $x\left(t^{i+1}, a_j(x^* + \Delta x^*)\right)$ and $d^n x\left(t^{i+1}, a_j(x^* + \Delta x^*)\right)/dt^{i+1^n}$, respectively at $t^{i+1} = t^{i+1}{}_0$. This can be concluded using the following equation based on the definition of super function:

$$\lim_{\Delta t \to 0} x\left(t^i, a_j(x^*)\right) = x\left(t^1{}_0, a_j(k_1)\right) \tag{3}$$

According to the provided definition here, the variation of $x^*$ can be selected so that it covers the variation of a nonlinear differential equation. As a result, a super function can be the answer of a nonlinear differential equation

### 1.4.1 Continuity of a super function

Consider a specified curve on the surface $x\left(t^i(t), a_j(x^*)\right)$ (i.e. $x^*$ is independent of $x$) in the three dimensional coordinates system $x - x^* - t$. The geometrical interpretation of a super function is the projection of this curve on the plane of $x - t$. Thus, the super function is continuous if the surface $x\left(t^i(t), a_j(x^*)\right)$ is continuous in the path of curve. The details of continuity conditions for a surface can be found in [1].

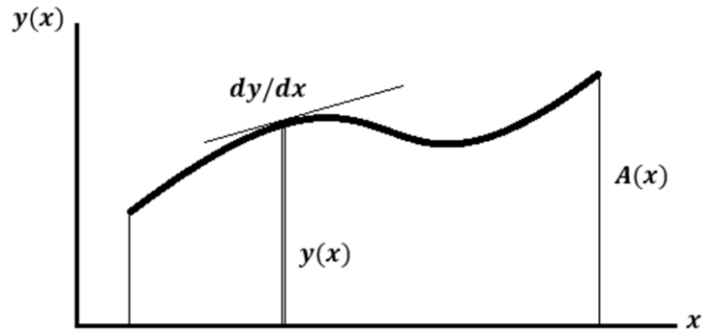

Figure 1. the nonlinear tem of a differential equation can be converted in a constant coefficient by tending the variation of governing equation to zero

Now, let us assume that we are to find the area of a rectangle. Consider Fig. 1; in this case $y = K$ and the area is obtained using $A = K(x_1 - x_0)$. However, this is not true for a variable $y(x)$, because there is no constant number to multiply it by $(x_1 - x_0)$ and for an arbitrary constant $x$ there is a corresponding constant $y$. To overcome the problem, we tend the domain of $(x_1 - x_0)$ to zero so that a constant $y$ can be found and in this case we are allowed to implement the multiplication operator that is $dA = y\, dx$ [1]. This means that the derivative of area with respect to the $x$ is $y$, Thus the area is obtained by integration. Now, the author relates this concept to the solution of a nonlinear differential equation. Regarding the presence of the nonlinear term in a variable differential equation, one can find a constant local differential equation by approaching its variation to zero which will be discussed soon.

**Theorem:** *Consider a nonlinear differential equation by the form $L(\ddot{x}, \dot{x}, y(x), a_j)$. The answer of this nonlinear differential equation is a super function $x\left(t^i, a_1 u(x^*)\right)$ in which, $u(x) = dy/dx$ and the local constant $x^*$ in each local function varies so that the equality of $x = x^*$ is satisfied for all points.*

**Proof:** As mentioned, the nonlinear term in a nonlinear differential equation gives different coefficients for different amplitudes. To find constant coefficient, the variation of $\Delta x$ should be approached to zero. Since $x = x(t)$, this can be applied by tending the variation of differential equation with respect to time to zero. Hence, one can write:

$$\frac{dL(\ddot{x}, \dot{x}, y(x), a_j)}{dt} = H(\dddot{x}, \ddot{x}, u(x^*)\dot{x}, a_j) \tag{4}$$

Since the variation of differential equation is tended to zero, Eq. (4) gives us a local constant linear differential equation with the time scale $t^i$ and $u(x^*)$ plays the role of a constant coefficient. Therefore, if the equality of $x = x^*$ is satisfied for all moments, the variation of answer is according to that of differential equation and this causes the nonlinear differential equation to be satisfied in any arbitrary time. Here, $H$ is the local differential equation for $x^* = k_1$ and $t^i = t^1$ and the corresponding local answer is $x(t^1, a_1 u(k_1))$.

## 2. The solution of pendulum equation

The governing equation of pendulum with arbitrary amplitude and initial conditions is as follows [2]:

$$\ddot{\theta} + \omega^2 \sin\theta = 0 \qquad (5)$$

The above equation is a nonlinear differential equation with a variable identity. One can replace Eq. (5) in (4) to obtain the local differential equation in time scale $t^i$:

$$\ddot{\theta} + \omega^2 \cos\theta^* \, \dot{\theta} = 0 \qquad (6)$$

By integrating Eq. (6) on $t^i$ and finding the integration constant using Taylor expansion of Eq. (5) we obtain the following equation:

$$\ddot{\theta} + \omega^2 \cos\theta^* \, \theta = \omega^2(\theta^* \cos\theta^* - \sin\theta^*) \qquad (7)$$

The general solution of Eq. (7) is extracted as follows and the angular velocity can be achieved by differentiating with respect to local time:

$$\theta\left(t^i, a_j(\theta^*)\right) = C^i{}_1 \sin\left(\sqrt{\omega^2 \cos\theta^*}\, t^i\right) + C^i{}_2 \cos\left(\sqrt{\omega^2 \cos\theta^*}\, t^i\right) + (\theta^* - \tan\theta^*) \qquad (8)$$

$$\dot{\theta}\left(t^i, a_j(\theta^*)\right) = \sqrt{\omega^2 \cos\theta^*}\left(C^i{}_1 \cos\left(\sqrt{\omega^2 \cos\theta^*}\, t^i\right) - C^i{}_2 \sin\left(\sqrt{\omega^2 \cos\theta^*}\, t^i\right)\right) \qquad (9)$$

Consider Eq. (7), it is clear that the amplitude is $-\pi \leq \theta \leq \pi$, so the answer is discrete in points $\theta = \pm\pi/2$ where it changes from harmonic form to exponential form. Therefore, the super function of answer for the amplitudes $|\theta| > \pi/2$ is obtained as:

$$\theta\left(t^i, a_j(\theta^*)\right) = C^i{}_1 e^{\sqrt{\omega^2|\cos\theta^*|}\, t^i} + C^i{}_2 e^{-\sqrt{\omega^2|\cos\theta^*|}\, t^i} + (\theta^* - \tan\theta^*) \qquad (10)$$

$$\dot{\theta}\left(t^i, a_j(\theta^*)\right) = \sqrt{\omega^2|\cos\theta^*|}\left(C^i{}_1 e^{\sqrt{\omega^2|\cos\theta^*|}\, t^i} - C^i{}_2 e^{-\sqrt{\omega^2|\cos\theta^*|}\, t^i}\right) \qquad (11)$$

If we consider $\theta^*$ as a completely independent variable, Eq. (8) and (10) represent a surface in which $C^i{}_1$ and $C^i{}_2$ should be obtained from infinite initial conditions $\theta^i = \theta^i{}_0$ and $\dot{\theta}^i = \dot{\theta}^i{}_0$, while the only available initial conditions are $\theta^0 = \theta^0{}_0$ and $\dot{\theta}^0 = \dot{\theta}^0{}_0$ in $t^0{}_0$ (i.e. $t^i = t$). To plot the super function numerically, the unknown constants of the local answer (i.e. $C^0{}_1$ and $C^0{}_2$), active in the first step, are determined using initial conditions. Now, the value of super function and its local derivative (i.e. angular velocity) are obtained at $t_0 + \Delta t$ using $\theta(t^0 + \Delta t^0, \theta^0{}_0)$. Therefore we have $\theta^1 = \theta^1{}_0$ and $\dot{\theta}^1 = \dot{\theta}^1{}_0$ in this point and these are initial conditions for time scale of $t^1$. By obtaining $C^1{}_1$ and $C^1{}_2$, the second active local answer is determined and one can obtain the value of super function in the second step using $\theta(t^1 + \Delta t^1, \theta^1{}_0)$. Applying the same manner, we can plot the super function for all points to satisfy governing equation and initial conditions. As it is seen, the general solution can be the projection of an arbitrary curve on the surfaces of Eq. (8) and (10) on the plane of $x - t$ and the particular solution is a definite curve depending on initial conditions. The natural frequency is found to be a variable in term of amplitude (i.e. local constant) $\sqrt{\omega^2 \cos \theta^*}$ and its average value can be calculated for a constant amplitude of motion.

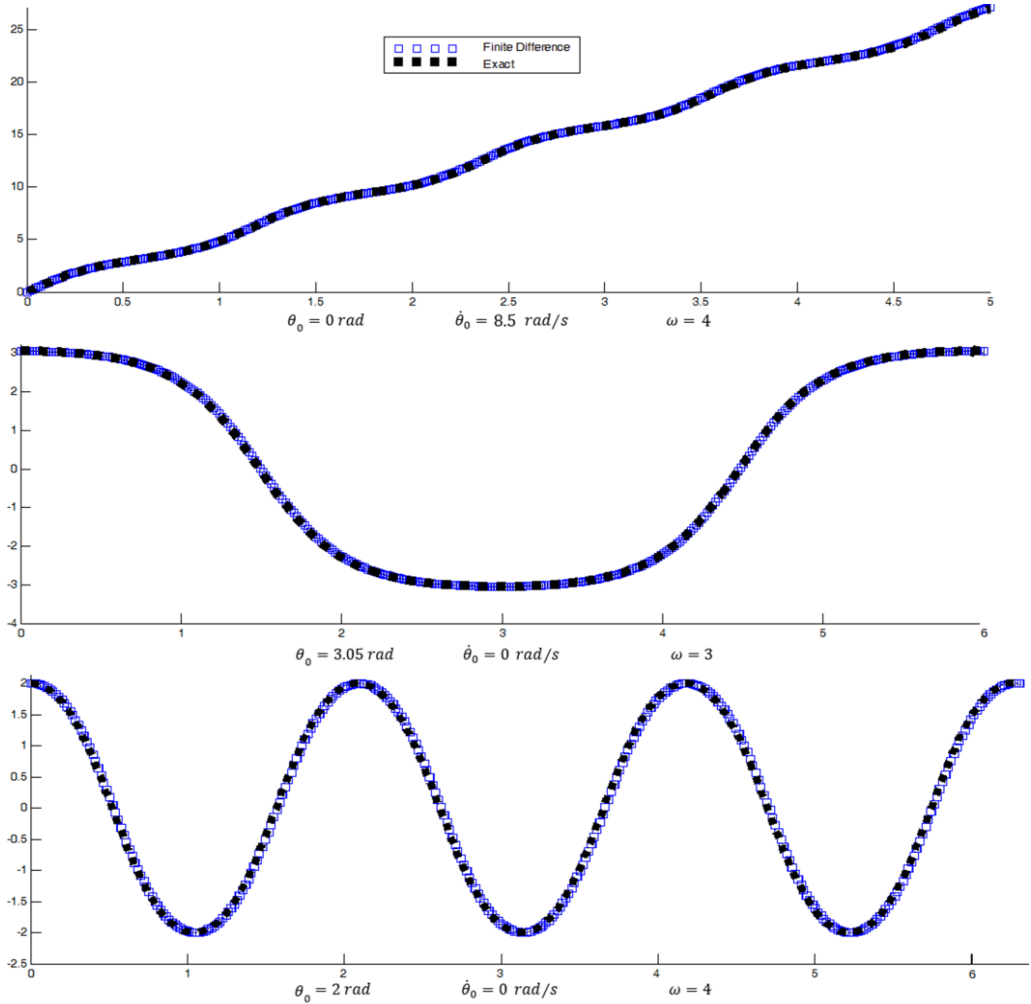

Figure 2. The results of new method are in agreement with those of finite difference

Using the explained algorithm above, the super function is modeled using a computer code in MATLAB to compare the results with those obtained from finite difference method. The results are shown in Fig. 2 and it is observed that there is an exact agreement for different initial conditions and any arbitrary amplitude of vibration.

**References:**

1. Jeffrey, A.: Advanced Engineering Mathematics. Elsevier. (2001)
2. Nayfeh, A.H., Mook, D.T.: Nonlinear Oscillation. John Wiley & Sons, Inc. (1995)